\documentclass[a4paper,12pt]{article}

\usepackage{graphics, epsfig}

\begin{document}

\author{C. Barrab\`es\thanks{E-mail : barrabes@celfi.phys.univ-tours.fr}\\     
\small Laboratoire de Math\'ematiques et Physique Th\'eorique\\
\small  CNRS/UPRES-A 6083, Universit\'e F. Rabelais, 37200 TOURS, France\\
P.A. Hogan\thanks{E-mail : phogan@ollamh.ucd.ie}\\
\small Mathematical Physics Department\\
\small  National University of Ireland Dublin, Belfield, Dublin 4, Ireland}

\title{A Class of Collisions of Plane Impulsive 
Light--Like Signals in General Relativity}
\date{}
\maketitle

\begin{abstract}
We present a systematic study of collisions of homogeneous, 
plane--fronted, impulsive light--like signals which do not interact 
after head--on collision. For the head--on collision of two such signals, 
six real parameters are involved, three from each of the incoming signals. 
We find two necessary conditions to be satisfied by these six parameters 
for the signals to be non--interacting after collision. We then solve 
the collision problem in general when these necessary conditions hold. 
After collision the two signals focus each other at Weyl curvature singularities 
on each others signal front. Our family of solutions contains some known 
collision solutions as special cases.
\end{abstract}
\thispagestyle{empty}
\newpage

\section{Introduction}\indent
A few models in general relativity of the head--on collision of 
homogeneous, plane--fronted, impulsive light--like signals are 
known (see below) in which the signals do not interact after collision. 
This is in stark contrast to what happens after the head--on 
collision of homogeneous, plane--fronted, impulsive gravitational 
waves \cite{KP} in which back--scattered gravitational radiation 
appears after the collision and a curvature singularity develops. 
In this paper we make a systematic study of collisions of homogeneous, 
plane--fronted, impulsive light--like signals which do not interact 
after head--on collision. In such cases the region of the space--
time model after collision is a portion of Minkowskian space--time 
and thus curvature singularity--free.

The space--time model of a general plane--fronted, impulsive, light--
like signal is described by a line--element involving one function 
of three variables. Such a signal in general consists of a plane--
fronted, light--like shell of matter accompanied by a plane--fronted, 
impulsive gravitational wave propagating through flat space--time. 
When specialised to be homogeneous this impulsive, light--like signal 
is characterised by three real parameters. One of these parameters is 
proportional to the relative energy density of the light--like shell of 
matter in the signal and the other two describe the two degrees of 
freedom of polarisation in the impulsive gravitational wave part of 
the signal. Thus for the head--on collision of two such signals, six 
real parameters are involved, three from each of the incoming signals. 
We find two necessary conditions to be satisfied by these six parameters 
for the signals to be non--interacting after collision. We then solve 
the collision problem in general when these necessary conditions hold. 
After collision the two signals focus each other at Weyl curvature singularities 
on each others signal front. Our family of solutions contains as 
special cases solutions found by Stoyanov \cite{ST}, Babala \cite{BA} 
and Feinstein and Senovilla \cite{FS}. 
 
The paper is organised as follows: In section 2 the general plane--fronted, 
impulsive, light--like signal propagating through flat space--time and 
determined by a single function of three variables is described and 
specialised to the homogeneous case. In section 3 we set up the head--on 
collision problem, derive the two necessary conditions on the six parameters 
involved for no interaction between the signals after collision and then 
solve Einstein's vacuum field equations for the region of space--time 
after collision. Finally in section 4 properties of the family of collision 
solutions derived in section 3 are studied and contact is made with the 
known solutions mentioned above.

\setcounter{equation}{0}
\section{Plane Impulsive Light--Like Signal}\indent
The general plane--fronted impulsive light--like signal 
propagating through Minkowskian space--time can easily be constructed. Its 
history in otherwise flat space--time $M$ is a null hyperplane 
$\cal N$ on which the Riemann tensor has a Dirac delta function 
singularity and it incorporates in general a plane impulsive gravitational 
wave and a plane, light--like shell of matter having relative energy density, isotropic 
pressure and anisotropic stress. In coordinates covering 
both sides of $\cal N$ the line--element of the space--time $M$ 
is given by \cite{BH1} 
\begin{equation}\label{2.1}
ds^2=-\left (dx+\frac{u_+}{F_v}\,dF_x\right )^2-\left (
dy+\frac{u_+}{F_v}\,dF_y\right )^2+2\,dv\,\left (du-\frac{u_+}{F_v}\,
dF_v\right )\ .
\end{equation}
Here $u_+=u\,\vartheta (u)$ with $\vartheta (u)$ the Heaviside step function 
which is equal to unity if $u>0$ and is equal to zero if $u<0$. The 
equation of $\cal N$ is $u=0$ and we have $u>0$ to the future 
of $\cal N$ while $u<0$ to the past of $\cal N$. The null geodesic integral 
curves of the vector field $\partial /\partial v$ generate $u=0$ (and 
all other null hyperplanes $u={\rm constant}$) and $v$ is an affine 
parameter along these generators. The coordinates $\xi ^a=(x, y, v)$, 
with $a=1, 2, 3$, are intrinsic to the null hyperplanes $u={\rm constant}$. Also 
$F=F(x, y, v)$ with partial derivatives indicated by subscripts and 
with $F_v\neq 0$. Calculation of the surface stress--energy tensor 
concentrated on $u=0$ shows that it describes light--like matter 
having relative energy density $\epsilon$, isotropic pressure $P$ 
and anisotropic stress $\Pi ^{ab}$ given by \cite{BH1}
\begin{eqnarray}\label{2.2}
\epsilon & = & -\frac{1}{8\pi}\,\frac{(F_{xx}+F_{yy})}{F_v}\ , \\
P & = & -\frac{1}{8\pi}\,\frac{F_{vv}}{F_v}\ , \\
\Pi ^{12} & = & -\frac{1}{8\pi}\,\frac{F_{xv}}{F_v}\ ,\qquad 
\Pi ^{13}=-\frac{1}{8\pi}\,\frac{F_{yv}}{F_v}\ ,
\end{eqnarray}
with all other components of $\Pi ^{ab}=\Pi ^{ba}$ vanishing. If any of these 
quantities are non--vanishing then the signal with history $\cal N$ 
has a part which is a light--like shell of matter. The coefficient of 
$\delta (u)$ in the Weyl tensor of $M$ has a Petrov type N 
part, with degenerate principal null direction $\partial /\partial v$, given 
in the Newman--Penrose notation by
\begin{equation}\label{2.5}
\hat\Psi _4=\frac{1}{2\,F_v}\,\left (F_{xx}-F_{yy}-2i\,F_{xy}\right )\ .
\end{equation}
If this is non--vanishing then the signal with history $\cal N$ has a 
part which is an impulsive gravitational wave. There is also a possibly 
non--zero Petrov type II coefficient of $\delta (u)$ in the Weyl tensor (
which if non--zero is associated with the presence of the light--like shell 
of matter described by (2.2)--(2.4)) given in Newman--Penrose notation 
by
\begin{equation}\label{2.6}
\hat\Psi _2=\frac{1}{3}\,\frac{F_{vv}}{F_v}\ ,\qquad\hat\Psi _3
=\frac{1}{\sqrt{2}\,F_v}\,\left (F_{xv}-iF_{yv}\right )\ .
\end{equation}
We see that if the surface stress--energy of the light--like shell is 
isotropic in the sense that $P$ and $\Pi^{ab}$ in (2.3), (2.4) vanish, then 
the quantities in (2.6) vanish. The acid test therefore for the existence 
of a light--like shell part of the signal with history $\cal N$ is the 
non--vanishing of some of (2.2)--(2.4), whereas the acid test for the existence 
of an impulsive gravitational wave part of the signal is the non--vanishing 
of (2.5). The calculation of (2.2)--(2.4) and of (2.5) involves independent sets 
of components of the jump in the transverse extrinsic curvature 3--tensor across 
$\cal N$ (see \cite{BI}). This fact allows the decomposition of an impulsive 
light--like signal into the two distinct parts to be carried out unambiguously 
\cite{BH2} \cite{BH3} in the case of {\it any} impulsive light--like 
signal propagating through {\it any} space--time and not just in the 
case of the plane--fronted impulsive light--like signal propagating 
through Minkowskian space--time under consideration here.

The null geodesic integral curves of $\partial/\partial u$ intersect the 
history $\cal N$ of the impulsive light--like signal. The Newman--Penrose 
spin coefficients $\sigma$ (the complex shear), $\rho$ (the real expansion), 
$\tau$ and $\gamma$ associated with this null geodesic congruence undergo 
the following jumps (indicated by square brackets) across $\cal N$:
\begin{equation}\label{2.7}
[\sigma ]=-\hat\Psi ^*_4\ ,\ [\rho ]=4\pi\,\epsilon\ ,\ 
[\tau ]=4\sqrt{2}\,\pi\,\left (\Pi ^{12}+i\Pi ^{13}\right )\ ,\ 
[\gamma ]=4\pi\,P\ ,
\end{equation}
with $\hat\Psi ^*_4$ the complex conjugate of $\hat\Psi _4$ in (\ref{2.5}) 
and $\epsilon\ ,\Pi^{12}\ , \Pi^{13}\ ,P$ given in (2.2)--(2.4). 
We see Penrose's observation \cite{P} that a jump in the shear of the null 
geodesic congruence transverse to $\cal N$ is necessary for the existence 
of a gravitational wave part to the signal and a jump in the expansion of 
this congruence is necessary for the shell part to have non--zero relative 
energy density.

The space--time $M$ with line--element (\ref{2.1}) can be constructed using a 
cut and paste technique in the style of Penrose \cite{P} as follows: The line--
element (2.1) with $u>0$ can be transformed to the manifestly flat form
\begin{equation}\label{2.8}
ds^2_+=-dx^2_+-dy^2_++2du_+dv_+\ ,
\end{equation}
by the transformation \cite{BH1}
\begin{eqnarray}\label{2.9}
x_+ & = & x+u\,\frac{F_x}{F_v}\ , \\
y_+ & = & y+u\,\frac{F_y}{F_v}\ , \\
v_+ & = & F+\frac{u}{2\,F_v}\,\left (F^2_x+F^2_y\right )\ ,\\
u_+ & = & \frac{u}{F_v}\ .
\end{eqnarray}
Thus (2.8) is the line--element of $M^+(u\geq 0)$. The line--element (2.1) 
for $u<0$ can be transformed trivially into flat form
\begin{equation}\label{2.13}
ds^2_-=-dx^2_--dy^2_-+2\,du_-dv_-\ ,
\end{equation}
by the tranformation
\begin{equation}\label{2.14}
x_-=x\ ,\qquad y_-=y\ ,\qquad v_-=v\ ,\qquad u_-=u\ .
\end{equation}
Thus (2.13) is the line--element of $M^-(u\leq 0)$. We see from (2.9)--(2.12) 
and (2.14) that on ${\cal N}(u=0)$ (which is clearly a null hyperplane), 
\begin{equation}\label{2.15}
x_+=x_- ,\qquad y_+=y_-\ ,\qquad v_+=F(x_-, y_-, v_-)\ ,
\end{equation}
and these matching conditions leave the induced line--element on ${\cal N}(u=0)$ 
invariant:
\begin{equation}\label{2.16}
dx^2_++dy^2_+=dx^2_-+dy^2_-\ .
\end{equation}
We have thus subdivided the space--time $M$ into two halves $M^+(u\geq 0)$ and 
$M^-(u\leq 0)$, each with boundary ${\cal N}(u=0)$, and we have then reattached 
the halves on $\cal N$ with the mapping (2.15) which preserves the induced line--
element on $\cal N$. 

For a homogeneous signal we take
\begin{equation}\label{2.17}
F(x, y, v)=v-\frac{a}{2}\,(x^2+y^2)+\frac{b}{2}\,(x^2-y^2)+c\,x\,y\ ,
\end{equation}
with $a, b, c$ constants and $a\geq 0$. This signal has an isotropic stress--energy 
($P=\Pi^{12}=\Pi^{13}=0$) with relative energy density (2.2) taking the form
\begin{equation}\label{2.18}
\epsilon =\frac{a}{4\pi}\ ,
\end{equation}
and is accompanied by an impulsive gravitational wave with 
\begin{equation}\label{2.19}
\hat\Psi _4=b-i\,c\ ,
\end{equation}
having the maximum two degrees of freedom of polarisation. The line--element 
(\ref{2.1}) reduces in this case to
\begin{eqnarray}\label{2.20}
ds^2 & = & -\left\{(1-(a-b)\,u_+)\,dx+c\,u_+dy\right\}^2-\left\{c\,u_+dx+(1-(a+b)\,u_+)
\,dy\right\}^2\nonumber\\
     &   & +2du\,dv\ .
\end{eqnarray}
This line--element has the Rosen--Szekeres form
\begin{eqnarray}\label{2.21}
ds^2 & = & -{\rm e}^{-U}\left ({\rm e}^V\cosh W\,dx^2-2\sinh W\,dx\,dy+{\rm e}^{-V}
\cosh W\,dy^2\right )\nonumber\\
     &   & +2\,{\rm e}^{-M}du\,dv\ ,
\end{eqnarray}
with
\begin{eqnarray}\label{2.22}
{\rm e}^{-U} & = & 1-2a\,u_++(a^2-b^2-c^2)\,u_+^2\ ,\\
{\rm e}^V & = & \left [\frac{(1-(a-b)\,u_+)^2+c^2u_+^2}{(1-(a+b)\,u_+)^2+c^2u_+^2}
\right ]^{\frac{1}{2}}\ ,\\
\sinh W & = & \frac{-2c\,u_+(1-a\,u_+)}{1-2a\,u_++(a^2-b^2-c^2)\,u_+^2}\ ,\\
M & = & 0\ .
\end{eqnarray}
In this paper we shall consider the head--on collision of this plane--fronted 
light--like signal with a signal of similar type. This latter signal is described by 
a space--time with line--element (2.21) with
\begin{eqnarray}\label{2.26}
{\rm e}^{-U} & = & 1-2\alpha\,v_++(\alpha ^2-\beta ^2-\gamma ^2)\,v_+^2\ ,\\
{\rm e}^V & = & \left [\frac{(1-(\alpha -\beta )\,v_+)^2+\gamma ^2v_+^2}{(1-(\alpha +\beta )
\,v_+)^2+\gamma ^2v_+^2}
\right ]^{\frac{1}{2}}\ ,\\
\sinh W & = & \frac{-2\gamma \,v_+(1-\alpha \,v_+)}{1-2\alpha \,v_++(\alpha ^2-\beta ^2-
\gamma ^2)\,v_+^2}\ ,\\
M & = & 0\ ,
\end{eqnarray}
where $\alpha , \beta , \gamma $ are real constants, $\alpha\geq 0$, and $v_+=v\,\vartheta (v)$. 
Here the history of the light--like signal is the null hyperplane $v=0$.

\setcounter{equation}{0}
\section{Interaction--Free Signals After Collision}\indent
The space--time model of the head--on collision of two plane--fronted, homogeneous, 
impulsive, light--like signals will have a line--element of the form (2.21) with, 
in general, $U, V, W$ and $M$ functions of $u, v$. In the region $v<0, u>0$ the 
functions $U, V, W$ and $M$ are given by (2.22)--(2.25) with $u_+=u$ and in the 
region $u<0, v>0$ these functions are given by (2.26)--(2.29) with $v_+=v$. The 
region $u<0, v<0$ has line--element (2.21) with $U=V=W=M=0$, which is consistent 
with (2.22)--(2.29) when both $u<0$ and $v<0$. The line--element in the region 
$u>0, v>0$ (after the collision) has the form (2.21) with $U, V, W$ and $M$ functions 
of $u, v$ satisfying the O'Brien--Synge \cite{OBS} junction conditions: For 
$v=0, u>0$ the functions $U, V, W$ and $M$ are given by (2.22)--(2.25) with 
$u_+=u$ and for $u=0, v>0$ the functions $U, V, W$ and $M$ are given by (2.26)--(2.29) 
with $v_+=v$. Einstein's vacuum field equations are to be solved for $U, V, W$ 
and $M$ in the region $u>0, v>0$ (after the collision) subject to these boundary 
(junction) conditions. These equations are \cite{GRIF} (with subscripts denoting 
partial derivatives)
\begin{eqnarray}
U_{uv}&=& U_u\,U_v\ ,\\\label{3.1}
2V_{uv}&=& U_u\,V_v+U_v\,V_u-2\left (V_u\,W_v+V_v\,W_u\right )
\tanh W\ ,\\\label{3.2}
2W_{uv}&=& U_u\,W_v+U_v\,W_u+2V_u\,V_v\,\sinh W\,\cosh W\ ,\\
\label{3.3}
2U_u\,M_u&=& -2U_{uu}+U_u^2+W_u^2+V_u^2\cosh ^2W\ ,\\\label{3.4}
2U_v\,M_v&=& -2U_{vv}+U_v^2+W_v^2+V_v^2\cosh ^2W\ ,\\\label{3.5}
2M_{uv}&=& -U_{uv}+W_u\,W_v+V_u\,V_v\,\cosh ^2W\ .
\end{eqnarray}
The first of these equations can immediately be solved \cite{EX} in conjuction 
with the boundary conditions to be satisfied by $U$ on $u=0, v>0$ 
and on $v=0, u>0$ to yield, in $u>0, v>0$,
\begin{equation}\label{3.7}
{\rm e}^{-U}=1-2\,a\,u-2\,\alpha\,v+(a^2-b^2-c^2)\,u^2+(\alpha ^2-\beta ^2-\gamma ^2)\,v^2\ .\end{equation}
The problem is to solve (3.2), (3.3) for $V, W$ 
subject to the boundary conditions and then to solve (3.4) 
and (3.5) for $M$. Equation (3.6) is the integrability 
condition for (3.4) and (3.5).

We can have non--interacting signals after collision by requiring 
that the region $u>0, v>0$ be, if possible, a portion of Minkowskian 
space--time. Necessary conditions for this to be possible can be obtained by 
examining approximate solutions of the vacuum field equations, 
subject to our boundary conditions, in the region $u>0, v>0$ near 
$u=0$ (i.e. for small $v>0$) and/or near $v=0$ (i.e. for small $u>0$). 
In this regard it is helpful to note that the non--identically vanishing 
components of the Riemann curvature tensor for the line--element (2.21) 
are given in Newman--Penrose notation by \cite{GRIF}
\begin{eqnarray}\label{3.8}
2\,\Psi _0 & = & B_v+\left (M_v-U_v\right )\,B+iB\,V_v\sinh W\ ,\\
2\,\Psi _2 & = & M_{uv}-\frac{1}{4}\,\left (A\,\bar B-\bar A\,B\right )\ ,\\
2\,\Psi _4^* & = & A_u+\left (M_u-U_u\right )\,A+iA\,V_u\sinh W\ ,
\end{eqnarray}
with 
\begin{equation}\label{3.11}
A=-V_u\cosh W+iW_u\ ,\qquad B=-V_v\cosh W+iW_v\ ,
\end{equation}
and the star, as before, denotes complex conjugation. Clearly necessary, 
but not necessarily sufficient, conditions for flatness in the region of 
space--time $u>0, v>0$ are
\begin{equation}\label{3.12}
M_{uv}=0\ ,
\end{equation}
and 
\begin{equation}\label{3.13}
A\,\bar B-\bar A\,B=0\ .
\end{equation}
The vacuum field equations (3.1) and (3.6) imply that (3.12) is equivalent to 
\begin{equation}\label{3.14}
A\,\bar B+\bar A\,B=2\,U_uU_v\ .
\end{equation}
The conditions (\ref{3.13}) and (\ref{3.14}) are to hold throughout the 
region $u>0, v>0$ and in particular near $u=0$ and/or $v=0$ in this region. 
It will be sufficient for our immediate purpose to calculate (3.13) and 
(3.14) at $u=v=0$ (strictly speaking in the limit $u\rightarrow 0^+$ 
and $v\rightarrow 0^+$). If {\it when} $v=0$ we put 
\begin{equation}\label{3.15}
A=S\,{\rm e}^{i\psi}\ ,\qquad B=R\,{\rm e}^{i\phi}\ ,
\end{equation}
we have immediately from our boundary conditions on $v=0\ (u>0)$
\begin{equation}\label{3.16}
S=\frac{2\,\sqrt{b^2+c^2}}{(1-a\,u)^2-(b^2+c^2)\,u^2}\ ,
\end{equation}
and $\psi =\theta +\pi$ with 
\begin{equation}\label{3.17}
{\rm e}^{i\theta}=\frac{b\{(1-a\,u)^2-(b^2+c^2)\,u^2\}+ic\,
\{(1-a\,u)^2+(b^2+c^2)\,u^2\}}{\sqrt{b^2+c^2}\,{\cal F}\,{\cal G}}\ ,
\end{equation}
with
\begin{equation}\label{3.18}
{\cal F}=\sqrt{(1-(a-b)\,u)^2+c^2u^2}\ ,\qquad {\cal G}=\sqrt{(1-(a+b)\,u)^2+c^2u^2} .
\end{equation} 
It is useful to note that
\begin{equation}\label{3.19}
\theta _u=\frac{4\,b\,c\,u\,(1-a\,u)}{{\cal F}^2{\cal G}^2}\ .
\end{equation}
The field equations (3.2) and (3.3) evaluated on $v=0$ give us the following 
differential equations for $R, \phi$:
\begin{eqnarray}\label{3.20}
R_u & = & \frac{\{a-(a^2-b^2-c^2)\,u\}\,R}{(1-a\,u)^2-(b^2+c^2)\,u^2}+\frac{2\,\alpha\,
\sqrt{b^2+c^2}\,\cos (\psi -\phi )}{\{(1-a\,u)^2-(b^2+c^2)\,u^2\}^2}
\ ,\\
\phi _u & = & \frac{4\,b\,c\,u\,(1-a\,u)}{{\cal F}^2{\cal G}^2}+\frac{2\,\alpha\,\sqrt{b^2+c^2}\,\sin (\psi -
\phi )}
{R\,\{(1-a\,u)^2-(b^2+c^2)\,u^2\}^2}\ .
\end{eqnarray}
Writing $\hat R=R\,\sqrt{(1-a\,u)^2-(b^2+c^2)\,u^2}$ and using (3.19) these can be 
simplified to
\begin{eqnarray}\label{3.22}
\hat R_u & = & \frac{2\,\alpha\,\sqrt{b^2+c^2}\,\cos (\psi -\phi )}
{\{(1-a\,u)^2-(b^2+c^2)\,u^2\}^{\frac{3}{2}}}\ ,\\
(\psi -\phi )_u & = & \frac{-2\,\alpha\,\sqrt{b^2+c^2}\,\sin (\psi -
\phi )}
{\hat R\,\{(1-a\,u)^2-(b^2+c^2)\,u^2\}^{\frac{3}{2}}}\ .
\end{eqnarray}
These give us
\begin{eqnarray}\label{3.24}
\hat R\,\sin (\psi -\phi ) & = & K\ ,\\
\hat R\cos (\psi -\phi ) & = & 
\frac{2\,\alpha\,\{a-(a^2-b^2-c^2)\,u\}}{\sqrt{b^2+c^2}\,\sqrt{(1-a\,u)^2-(b^2+c^2)\,u^2}}
+C\ ,
\end{eqnarray}
where $K, C$ are constants of integration. From our boundary conditions we have, when $u=v=0$, 
\begin{equation}\label{3.25}
{\rm e}^{i\psi}=-\frac{(b+ic)}{a}\ ,\ {\rm e}^{i\phi}=-\frac{(\beta +i\gamma )}{\alpha}
\ ,\ R=2\,\alpha\ .
\end{equation}
This determines the constants $K, C$ and writing $\hat R$ in terms of $R$ we have 
from (3.24) and (3.25)
\begin{eqnarray}\label{3.26}
R\,\sin (\psi -\phi ) & = & \frac{2\,(\beta\,c-\gamma\,b)}{\sqrt{b^2+c^2}\,
\sqrt{(1-a\,u)^2-(b^2+c^2)\,u^2}}\ ,\\
R\,\cos (\psi -\phi ) & = & \frac{2\,\alpha\,\{a-(a^2-b^2-c^2)\,u\}}
{\sqrt{b^2+c^2}\,\{(1-a\,u)^2-(b^2+c^2)\,u^2\}}\nonumber\\ 
& & +\frac{2\,(\beta\,b +\gamma\,c -\alpha\,a )}
{\sqrt{b^2+c^2}\,\sqrt{(1-a\,u)^2-(b^2+c^2)\,u^2}}\ .
\end{eqnarray}
Now (3.13) evaluated at $v=0$ gives
\begin{equation}\label{3.29}
\sin (\psi -\phi )=0\ ,
\end{equation}
and by (3.27) this is equivalent to 
\begin{equation}\label{3.30}
\beta\,c-\gamma\,b=0\ .
\end{equation}
Next using $U$ given by (3.7), $S$ given by (3.16) and $R\,\cos (\psi -\phi )$ 
by (3.28), the equation (3.14) evaluated at $v=0$ is found to be equivalent to
\begin{equation}\label{3.31}
\beta\,b+\gamma\,c-\alpha\,a=0\ .
\end{equation}
Hence {\it the relations (3.30) and (3.31)} between the three real constants 
$a, b, c$ associated with one in--coming light--like signal and the three 
real constants $\alpha , \beta , \gamma$ associated with the second incoming 
light--like signal {\it are necessary conditions for the region of space--time 
$u>0, v>0$ (after the collision) to be flat}. We note that (3.30) and (3.31) 
are invariant under the interchange of $a, b, c$ with $\alpha , \beta , \gamma$ 
respectively and so would also emerge had we evaluated (3.13) and (3.14) at 
$u=0$ instead of at $v=0$. We note that (3.30) and (3.31) exclude the case of 
a collision of two plane impulsive gravitational waves \cite{KP} (corresponding 
to $b\neq 0, \beta\neq 0, a=\alpha =\gamma =c=0$) and also a collision of two 
plane light--like shells \cite{DT} (corresponding to $a\neq 0, \alpha\neq 0, 
b=c=\beta =\gamma =0$). In those two cases the spacetime after the collision
is not flat and contains a curvature singularity. 

We shall henceforth assume that the conditions (3.30) and (3.31) are satisfied 
and we proceed to solve the vacuum field equations (3.2)--(3.6) in the region 
$u>0, v>0$ with the boundary conditions on $U, V, W$ and $M$ indicated in the 
opening paragraph of this section. We use a procedure described in 
\cite{PTP} to which we must refer the reader for details. The solution derived 
can of course be verified by substitution into the field equations. The preliminary 
calculations necessary to implement the procedure in \cite{PTP} have been 
carried out above. These are the calculations determining $S, \psi , R, 
\phi$ in (3.15). The results are given by (3.16), (3.17), (3.27) and (3.28) 
with (3.30) and (3.31) now holding. Using these as starting point we find 
that, in addition to $U$ already given in $u>0, v>0$ by (3.7), we obtain
\begin{eqnarray}\label{3.32}
{\rm e}^V & = & \left [\frac{\{1-(a-b)\,u-(\alpha -\beta )\,v\}^2+(c\,u+\gamma\,v)^2}
{\{1-(a+b)\,u-(\alpha +\beta )\,v\}^2+(c\,u+\gamma\,v)^2}\right ]^{\frac{1}{2}}\ ,\\
\sinh W & = & -2(c\,u+\gamma\,v)\,(1-a\,u-\alpha\,v)\,{\rm e}^U\ ,\\
M & = & 0\ .
\end{eqnarray}
If in these functions we replace $u$ by $u_+=u\,\vartheta (u)$ and $v$ by $v_+
=v\,\vartheta (v)$ then we can include in a single expression in each case the form of 
the function in the four different regions of the collision space--time. Having 
done this and having substituted the functions into the line--element (2.21) 
we arrive at the final form of our line--element:
\begin{eqnarray}\label{3.35}
ds^2 & = & -\{(1-(a-b)\,u_+-(\alpha -\beta )\,v_+)\,dx+(c\,u_++\gamma\,v_+)\,dy\}^2
\nonumber\\
& &  -\{(c\,u_++\gamma\,v_+)\,dx+(1-(a+b)\,u_+-(\alpha +\beta )\,v_+)\,dy\}^2\nonumber\\
& & +2\,du\,dv\ ,
\end{eqnarray}
with the six parameters $a, b, c, \alpha , \beta , \gamma$ satisfying (3.30) and 
(3.31). The non--identically vanishing Weyl tensor components for the space--time 
with line--element (3.35) are, in Newman--Penrose notation,
\begin{eqnarray}\label{3.36}
\Psi _0 & = & -\frac{\alpha\,\{a-(a^2-b^2-c^2)\,u_+\}\,{\cal P}}
{(b^2+c^2)\,{\cal F}\,{\cal G}\,{\cal P}_1}\,\delta (v)\ ,\\
\Psi _4 & = & -\frac{a\,\{\alpha -(\alpha ^2-\beta ^2-\gamma ^2)\,v_+\}\,{\cal Q}}
{(\beta ^2+\gamma ^2)\,{\cal F}'\,{\cal G}'\,{\cal Q}_1}
\,\delta (u)\ .
\end{eqnarray}
Here 
\begin{eqnarray}\label{3.38}
{\cal P} & = & b\,{\cal P}_1+i\,c\,{\cal P}_2\ ,\\
{\cal Q} & = & \beta
\,{\cal Q}_1-i\,\gamma \,{\cal Q}_2\ ,\\
{\cal P}_1 & = & \{(1-a\,u_+)^2-(b^2+c^2)\,u_+^2\}\ ,\\
{\cal P}_2 & = & \{(1-a\,u_+)^2+(b^2+c^2)\,u_+^2\}\ ,\\
{\cal Q}_1 & = & \{(1-\alpha \,v_+)^2-(\beta ^2+\gamma ^2)\,v_+^2\}\ ,\\
{\cal Q}_2 & = & \{(1-\alpha \,v_+)^2+(\beta ^2+\gamma ^2)\,v_+^2\}\ ,\\
{\cal F}' & = & \sqrt{(1-(\alpha -\beta )\,v_+)^2+\gamma ^2v_+^2}\ ,\\ 
{\cal G}' & = & \sqrt{(1-(\alpha +\beta )\,v_+)^2+\gamma ^2v_+^2}\ ,
\end{eqnarray}
and ${\cal F}$, ${\cal G}$ are given by (3.18) with $u$ replaced by $u_+$. Labelling 
the coordinates $x^{\mu}=(x, y, u, v)$ with $\mu =1, 2, 3, 4$ the Ricci tensor 
components for the space--time with line--element (3.35) are
\begin{equation}\label{3.46}
R_{\mu\nu}=-\frac{2\,\alpha}{{\cal P}_1}\,\delta (v)\,l_{\mu}\,l_{\nu}-
\frac{2\,a}{{\cal Q}_1}\,\delta (u)\,n_{\mu}\,n_{\nu}\ ,
\end{equation}
with $l^{\mu}=\delta ^{\mu}_3$ and $n^{\mu}=\delta ^{\mu}_4$. Finally we note that 
on $v=0$ the (real) expansion $\rho _l$ and the modulus of the complex 
shear $|\sigma _l|$ of the null 
geodesic integral curves of $l^\mu$ are given by
\begin{equation}\label{3.47}
\rho _l=\frac{a\,\vartheta (u)-(a^2-b^2-c^2)\,u_+}{{\cal P}_1}\ ,\qquad 
|\sigma _l|=\frac{\sqrt{b^2+c^2}\,\vartheta (u)}{{\cal P}_1}\ ,
\end{equation}
while on $u=0$ the (real) expansion $\rho _n$ and the modulus of the complex shear 
$|\sigma _n|$ of the null geodesic integral curves of $n^\mu$ are given by
\begin{equation}\label{3.48}
\rho _n=\frac{\alpha\,\vartheta (v)-(\alpha ^2-\beta ^2-\gamma ^2)\,v_+}{{\cal Q}_1}\ ,\qquad 
|\sigma _n|=\frac{\sqrt{\beta ^2+\gamma ^2}\,\vartheta (v)}{{\cal Q}_1}\ .
\end{equation}

\setcounter{equation}{0}
\section{Discussion}\indent
We remark that for the space--time with line--element (3.35) it follows from (3.36), 
(3.37) and (3.46) that the region $u>0, v>0$ after the collision is flat. To see this 
explicitly we consider two cases separately.

\noindent
${\rm I}\ (c=0=\gamma ):$ for $u>0, v>0$ we have
\begin{equation}\label{4.1}
ds^2=-\{1-(a-b)\,u-(\alpha -\beta )\,v\}^2dx^2-\{1-(a+b)\,u-(\alpha +\beta )\,v\}^2
dy^2+2\,du\,dv\ ,
\end{equation}
with $a\,\alpha -b\,\beta =0$. We first note that if $a=\pm b\ 
(\Leftrightarrow\alpha\pm\beta )$ then it is easy to transform (4.1) to 
Minkowskian form \cite{TRA} . Now assume $a\neq\pm b\ (\Leftrightarrow\alpha\neq\pm\beta )$ and 
make the transformation
\begin{eqnarray}\label{4.2}
u' & = & 1-(a-b)\,u-(\alpha -\beta )\,v\ ,\\
v' & = & 1-(a+b)\,u-(\alpha +\beta )\,v\ .
\end{eqnarray}
This results in (4.1) taking the form
\begin{equation}\label{4.4}
ds^2=-u'^2dx^2-v'^2dy^2-\lambda ^2du'^2+\mu ^2dv'^2\ ,
\end{equation}
with $\lambda ^{-2}=-2\,(\alpha -\beta )\,(a-b),\  \mu ^{-2}=
2\,(\alpha +\beta )\,(a+b)$. We note that $(\alpha -\beta )\,(a-b)<0$ and 
also $(\alpha ^2-\beta ^2)\,(a^2-b^2)=-(\alpha\,b-\beta\,a)^2$ and so $(\alpha +\beta )\,
(a+b)>0$. Then with $\bar u=\lambda\,u', \bar v=\mu\,v', \bar x=\lambda ^{-1}x, 
\bar y=\mu ^{-1}y$ we have
\begin{equation}\label{4.5}
ds^2=-\bar u^2d\bar x^2-\bar v^2d\bar y^2-d\bar u^2+d\bar v^2\ .
\end{equation}
Now put $X=\bar u\,\cos\bar x,\ Y=\bar u\,\sin\bar x,\ Z=\bar v\,\sinh\bar y,\ 
T=\bar v\cosh\bar y$ and we arrive at 
\begin{equation}\label{4.6}
ds^2=-dX^2-dY^2-dZ^2+dT^2\ .
\end{equation}
We note that $0\leq\bar x <2\,\pi$ and so $0\leq x <2\,\pi\,\lambda$. This periodicity 
in the coordinate $x$ has been noted already in a special case \cite{BA} of (3.35) which 
we draw attention to below.

\noindent
${\rm II} (c\neq 0\neq\gamma ):$ for $u>0, v>0$ in this general case we have from 
(3.35)
\begin{eqnarray}\label{4.7}
ds^2 & = & -\{(1-(a-b)\,u-(\alpha -\beta )\,v)\,dx+(c\,u+\gamma\,v)\,dy\}^2
\nonumber\\
& &  -\{(c\,u+\gamma\,v)\,dx+(1-(a+b)\,u-(\alpha +\beta )\,v)\,dy\}^2\nonumber\\
& & +2\,du\,dv\ ,
\end{eqnarray}
with $a\,\alpha -b\,\beta =c\,\gamma$ and $\beta\,c=b\,\gamma$. The rotation
\begin{equation}\label{4.8}
x=\frac{x'+\Lambda\,y'}{\sqrt{1+\Lambda ^2}}\ ,\qquad y=
\frac{-\Lambda\,x'+y'}{\sqrt{1+\Lambda ^2}}\ ,
\end{equation}
with $\Lambda$ satisfying
\begin{equation}\label{4.9}
\Lambda ^2-2\,\frac{b}{c}\,\Lambda -1=0\ ,
\end{equation}
results in (4.7) taking the form
\begin{equation}\label{4.10}
ds^2=-\{1-(a'-b')\,u-(\alpha '-\beta ')\,v\}^2dx'^2-\{1-(a'+b')\,u-(\alpha '+\beta ')\,v\}^2
dy'^2+2\,du\,dv\ ,
\end{equation}
with
\begin{equation}\label{4.11}
a'=a\ ,\qquad b'=b-\Lambda\,c\ ,\qquad \alpha '=\alpha\ ,\qquad\beta '=\beta -\Lambda\,\gamma\ ,
\end{equation}
from which it follows that
\begin{equation}\label{4.12}
a'\alpha '-b'\beta '=0\ .
\end{equation}
Now (4.10) with (4.12) is identical to the case I considered above. It thus 
follows that (4.10) (and hence (4.7)) can be written in manifestly Minkowskian 
form and that in this case $x'$ is a periodic coordinate. 

The special case of (3.35) corresponding to $a=b, \alpha =\beta$ and $c=\gamma =0$ 
gives a subset of a family of solutions originally found by Stoyanov \cite{ST}. The 
solution (3.35) in this case describes the collision of two linearly polarised 
plane impulsive gravitational waves each sharing its wave front with a plane 
light--like shell with relative energy density proportional to the amplitude 
of the wave. The special case of (3.35) corresponding to $a\neq 0, b=0, \alpha =0, 
\beta\neq 0, c=\gamma =0$, describes the collision of a linearly polarised 
plane impulsive gravitational wave with a plane light--like shell and was found by 
Babala \cite{BA}. The generalisation of this for which $a\neq 0, b=c=0, \alpha =0, 
\beta\neq 0, \gamma\neq 0$ was found by Feinstein and Senovilla \cite{FS}.

We see from (3.47) and (3.48) that the signals involved in the collision 
focus each other after collision on their signal fronts at $v=0, {\cal P}_1=0$ and at $u=0, 
{\cal Q}_1=0$, where it follows from (3.36) and (3.37) that the Weyl curvature 
is singular and from (3.46) that the surface stress--energy of the light--like 
shells (after collision) is singular in each case. On the signal front 
$u=0, v>0$ or $v=0, u>0$ the signals are focused on two cylinders with 
elliptic cross--sections in general. It is important to note in 
this regard that when $c=0$ we have ${\cal F}\,{\cal G}=0\ \Leftrightarrow\ {\cal P}_1=0$ 
and similarly when $\gamma =0$ we have ${\cal F}'\,{\cal G}'=0\ \Leftrightarrow\ {\cal Q}_1=0$. 
Also since the coordinates $x$ in case I above and $x'$ in case II are periodic 
the space--time region $u>0, v>0$ does not extend to infinity but fills a 
cylinder with, in general, elliptic cross--sections. A topological singularity occurs 
on $(a+\sqrt{b^2+c^2})u+(\alpha +\sqrt{\beta ^2+\gamma ^2})v=1$, and the space--
time cannot be uniquely extended beyond this singularity. For an exceptional choice of the parameters 
(see \cite{BA}, for example) the two cylinders on which the signals focus each 
other degenerate into a point and a circle.

\noindent
\section*{Acknowledgment}\noindent
This collaboration has been funded by the Minist\`ere des Affaires 
\'Etrang\`eres, D.C.R.I. 220/SUR/R.


\begin{thebibliography}{99}
\bibitem{EX} It is well known (see \cite{GRIF}) that (\ref{3.1}) results in ${\rm e}^{-U}=
f(u)+g(v)$ for some functions $f, g$ and these functions 
are immediately determined by the boundary values (\ref{2.22}) and 
(\ref{2.26}) of $U$.
\bibitem{TRA} If $a=b$ then $\alpha =\beta$ and 
\begin{eqnarray}
-dx^2-(1-2\,a\,u-2\,\alpha\,v)^2dy^2+2du\,dv & = & -dx^2-dY^2-dz^2+dt^2\ ,\nonumber
\end{eqnarray}
with $Y=\frac{(1-2\,a\,u-2\,\alpha\,v)}{\sqrt{8a\,\alpha}}\,\sinh\left (\sqrt{8a\,\alpha}\,y
\right ),\ z=\frac{1+2a\,u-2\alpha\,v}{\sqrt{8a\,\alpha}}$ and $t=
\frac{(1-2\,a\,u-2\,\alpha\,v)}{\sqrt{8a\,\alpha}}\,\cosh\left (\sqrt{8a\,\alpha}\,y
\right )$ and we note that for physical reasons $a>0$ and $\alpha >0$. The case $a=-b\  (
\Leftrightarrow \alpha =-\beta )$ is the same as this with $x$ and $y$ interchanged.
\end{thebibliography}
\end{document}